# ONE-DIMENSIONAL ULTRASOUND PROPAGATION AT STRATIFIED GAS: GROSS-JACKSON MODEL

M. Solovchuk[1], S. Leble[2]

[1]Theoretical Physics Department, I. Kant State University of Russia,
Al.Nevsky street 14, Kaliningrad , 236041, Russia, E-mail:solovchuk@yandex.ru

[2]Faculty of Technical Physics and Applied Mathematics, Gdansk University of Technology,
ul. G.Narutowicza 11/12, 80-952 Gdansk-Wrzeszcz Poland, E-mail: leble@mif.pg.gda.pl

The system of hydrodynamic-type equations, derived by two-side distribution function for a stratified gas in gravity field is applied to a problem of ultrasound. The theory is based on Gross-Jackson kinetic equation, which solution is built by means of locally equilibrium distribution function with different local parameters for molecules moving "up" and "down". The background state and linearized version of the obtained system is studied and compared with other results and experiments at arbitrary Knudsen numbers.

The problem of a generation by a moving plane in a rarefied gas is explored. The WKB solutions for ultrasound in a stratified medium are constructed in explicit form, evaluated and plotted.

## 1. Introduction

Recently the problems of Kn regime wave propagation was revisited in connection with general fluid mechanics and nonsingular perturbation method development [1-3]. Generalized Boltzmann theories [4-6] also contributed in a progress with respect to this important problem.

In [7] the propagation of one-dimension disturbance was studied on the base of the method of a piecewise continuous distribution function launched in a pioneering paper of Lees [8] and applied for a gas in gravity field in [1,9]. We derived hydrodynamic-type equations for a gas perturbations in gravity field so that the Knudsen number depends on the (vertical) coordinate. The generalization to three dimensions is given at [10,11].

In a recent papers[7,10] it was shown that for speed ratio and attenuation factor the BGK-model provides good agreement with experimental data[12,13]. Since the BGK-model has a Prandtl number 1, wich is incorrect for monatomic gases, a certain judicious choice of mean free path is always required. It is of interest here to examine the higher order kinetic model to assess the model dependence of the results. We have considered one such class of higher order models, the Gross-Jackson model[14], in this paper.

## 2. Generalized fluid dynamics equations

The derivation of the hydrodynamic-type equations is based on Gross-Jackson kinetic equation[14], that looks in one-dimensional case like:

$$\frac{\partial \varphi}{\partial t}+\vec{V}\cdot\frac{\partial \varphi}{\partial r}-g\frac{\partial \varphi}{\partial V_z}=\nu\cdot\left(\sum_{I=1}^{3}<\chi_I,\varphi>\varphi_I+\frac{1}{3}<\chi_4,\varphi>\chi_4-\varphi\right), \qquad (1)$$

here $\varphi(t,z,\vec{V})=(f-f_0)/f_0$ represents the perturbation of the distribution function from the Maxwellian $f_0$, $t$ is time, $\vec{V}$ is velocity of a particle of a gas, $\vec{r}$ is coordinate, $\nu(z)=\nu_0\exp(-z/H)$ is the effective frequency of collisions between particles of gas at height $z$, $H=kT/mg$ is a parameter of the gas stratification.

The moments of distribution function are defined by:

$$M_I =<\chi_I,\varphi>= \frac{1}{\pi^{3/2}}\int d\vec{c}\exp(-c^2)\chi_I(\vec{c})\varphi, \qquad (2)$$

where $\vec{C}=\vec{V}/V_T$ is the dimensionless velocity, $V_T=\sqrt{2kT/m}$ denotes the average thermal velocity of particles of gas. The first six eigen functions $\chi_I$ of the linearized collision operator are:

$$\chi_1=1, \qquad \chi_2=\sqrt{2}c_z, \qquad \chi_3=\sqrt{\frac{2}{3}}(\frac{3}{2}-c^2),$$

$$\chi_4=\sqrt{5}c_z(1-\frac{2}{5}c^2), \quad \chi_5=\frac{1}{\sqrt{3}}(c^2-3c_z^2), \quad \chi_6=\sqrt{\frac{6}{5}}c_z(c^2-\frac{5}{3}c_z^2), \qquad (3)$$

The expressions for the moments of distribution function are:

$$M_1=\frac{n-n_0}{n_0}, \quad M_2=\sqrt{2}\frac{n}{n_0}\frac{U_z}{V_T}, \quad M_3=-\sqrt{\frac{2}{3}}\frac{n}{n_0}\frac{U_z^2}{V_T^2}-\sqrt{\frac{3}{2}}\frac{n}{n_0}\frac{T-T_0}{T_0}$$

$$M_4=-\frac{3}{\sqrt{5}}\frac{n}{n_0}\frac{T-T_0}{T_0}\frac{U_z}{V_T}-\frac{2}{\sqrt{5}}\frac{P_{zz}-nkT_0}{n_0kT_0}\frac{U_z}{V_T}-\frac{2}{\sqrt{5}}\frac{n}{n_0}\frac{U_z^3}{V_T^3}-\frac{4}{\sqrt{5}}\frac{q_z}{mn_0V_T^3} \qquad (4)$$

$$M_5=\frac{1}{\sqrt{3}}(\frac{3}{2}\frac{nkT-P_{zz}}{n_0kT_0}-2\frac{n}{n_0}\frac{U_z^2}{V_T^2})$$

$$M_6=\sqrt{\frac{6}{5}}\left(-\frac{2}{3}\frac{n}{n_0}\frac{U_z^3}{V_T^3}+\frac{U_z}{V_T}\frac{3}{2}\frac{nkT-P_{zz}}{n_0kT_0}+\frac{2}{n_0V_T^3m}(q_z-\frac{5}{3}\bar{q}_z)\right)$$

In linear approach the three momenta are proportional to mass density $n$, velocity $U_z$ and temperature $T$ variations. Here $P_{zz}$ is the diagonal component of the pressure tensor, $q_z$ is a vertical component of a heat flux vector, $\bar{q}_z$ is a parameter having dimension of the heat flux.

Following the idea of the method of piecewise continuous distribution functions let's search for the solution $\varphi$ of the equations(1) as a combination of two locally equilibrium distribution functions, each of which gives the contribution in its own area of velocities space:

$$\varphi = \begin{cases} \varphi^+ = \dfrac{n^+-n_0}{n_0}+2C_z\dfrac{U_z^+}{V_T}+(C^2-\dfrac{3}{2})\dfrac{T^+-T_0}{T_0}, & V_z\geq 0 \\ \\ \varphi^- = \dfrac{n^--n_0}{n_0}+2C_z\dfrac{U_z^-}{V_T}+(C^2-\dfrac{3}{2})\dfrac{T^--T_0}{T_0}, & V_z<0 \end{cases} \qquad (5)$$

The increase of the number of parameters of distribution function results in that the distribution function differs from a local-equilibrium one and describes deviations from hydrodynamical regime. In the range of small Knudsen numbers $l \ll L$ we automatically have $n^+ = n^-, U^+ = U^-, T^+ = T^-$ and distribution function reproduces the hydrodynamics of Euler and at the small difference of the functional 'up' and 'down' parameters - the Navier-Stokes equations. In the range of big Knudsen numbers the theory gives solutions of collisionless problems [9].

The functional parameters are linked to the moments of distribution function:

$$\frac{n^\pm - n_0}{n_0} = \sqrt{3} M_5 + M_1 \pm \frac{3}{10}\sqrt{5\pi} M_4 \pm \frac{7\sqrt{30\pi}}{20} M_6$$

$$\frac{U_z^\pm}{V_T} = -\frac{1}{\sqrt{5}} M_4 \mp \frac{\sqrt{3\pi}}{2} M_5 - \frac{2}{5}\sqrt{30} M_6 + \frac{1}{\sqrt{2}} M_2$$

$$\frac{T^\pm - T_0}{T_0} = \frac{1}{\sqrt{3}} M_5 - \frac{\sqrt{6}}{3} M_3 \mp \frac{\sqrt{\pi}}{\sqrt{5}} M_4 \pm \frac{\sqrt{30\pi}}{20} M_6$$

If we now multiply Gross-Jackson kinetic equation by $\chi_I$ and integrate over velocity space, we obtain the fluid dynamic equations:

$$\frac{\partial}{\partial t} M_1 + \frac{V_T}{\sqrt{2}} \frac{\partial}{\partial z} M_2 - \frac{V_T}{\sqrt{2} H} M_2 = 0$$

$$\frac{\partial M_2}{\partial t} + V_T \frac{\partial}{\partial z}(\frac{1}{\sqrt{2}} M_1 - \frac{1}{\sqrt{3}} M_3 - \frac{\sqrt{2}}{\sqrt{3}} M_5) + \frac{V_T}{H}(\frac{1}{\sqrt{3}} M_3 + \frac{\sqrt{2}}{\sqrt{3}} M_5) = 0$$

$$\frac{\partial M_3}{\partial t} + V_T \frac{\partial}{\partial z}(-\frac{1}{\sqrt{3}} M_2 + \sqrt{\frac{5}{6}} M_4) - \sqrt{\frac{5}{6}} \frac{V_T}{H} M_4 = 0 \qquad (6)$$

$$\frac{\partial M_4}{\partial t} + V_T \frac{\partial}{\partial z}(\sqrt{\frac{5}{6}} M_3 + \frac{1}{30}\sqrt{15} M_5) + \frac{\sqrt{15}}{10} \frac{V_T}{H} M_5 = -\frac{2}{3} v M_4$$

$$\frac{\partial M_5}{\partial t} + V_T \frac{\partial}{\partial z}(-\sqrt{\frac{2}{3}} M_2 + \frac{2}{\sqrt{15}} M_4 + \frac{3}{\sqrt{10}} M_6) - \frac{V_T}{H}(\frac{2}{\sqrt{15}} M_4 + \frac{\sqrt{3}}{2} M_6) = -v M_5$$

$$\frac{\partial M_6}{\partial t} + \sqrt{\frac{2}{5}} V_T \frac{\partial}{\partial z} M_5 - \frac{V_T}{H} \frac{M_5}{\sqrt{10}} = -v M_6$$

Here we have used the expressions for the two integrals:

$$J_1 = <C_z^2 C^2 \varphi> = \frac{5}{4} M_1 - \frac{5}{\sqrt{6}} M_3 - \frac{11}{12}\sqrt{3} M_5$$

$$J_2 = <C_z^4 \varphi> = \frac{3}{4} M_1 - \frac{\sqrt{6}}{2} M_3 - \frac{3}{4}\sqrt{3} M_5$$

Hence a modification of the procedure for deriving fluid mechanics (hydrodynamic-type) equations from the kinetic theory is proposed, it generalizes the Navier-Stokes at arbitrary density (Knudsen numbers).

Our method gives a reasonable agreement with the experimental data in the case of homogeneous gas [7]. In the paper[7] the expressions for $J_{1,2}$ are obtained with account some nonlinear terms, that finally lead to more exact results in transition region.

## 3. A limiting case of gas oscillations at high frequencies of collisions (small knudsen numbers).

Let us consider a system in the hydrodynamical limit $\nu \to \infty$. It follows from the last three equations of the system (6) that $M_4, M_5, M_6 \ll M_1, M_2, M_3$. Next assume $\nu^{-1} = 0$ in the zero order by the parameter $\nu^{-1}$. We have $M_4, M_5, M_6 = 0$ and the system(6) tends to the linearized Euler's system:

$$M_{1t} + \frac{1}{\sqrt{2}} V_T M_{2Z} - \frac{1}{\sqrt{2}} \frac{V_T}{H} M_2 = 0$$

$$M_{2t} + V_T(\frac{1}{\sqrt{2}} M_{1Z} - \frac{1}{\sqrt{3}} M_{3Z}) + \frac{1}{\sqrt{3}} \frac{V_T}{H} M_3 = 0 \qquad (7)$$

$$M_{3t} - \frac{V_T}{\sqrt{3}} M_{2Z} = 0$$

The moments $M_4, M_5, M_6$ belong to the next order of the parameter $\nu^{-1}$. Then from the last three equations of the system (6), one obtains following relations

$$M_4 = -\frac{3}{2}\sqrt{\frac{5}{6}} \frac{V_T}{\nu} M_{3Z}, \quad M_5 = \sqrt{\frac{2}{3}} \frac{V_T}{\nu} M_{2Z}, \quad M_6 = 0, \qquad (8)$$

Further substituting (8) in the first three equations of the system (6) we obtain:

$$M_{1t} + \frac{1}{\sqrt{2}} V_T M_{2Z} - \frac{1}{\sqrt{2}} \frac{V_T}{H} M_2 = 0$$

$$M_{2t} + V_T(\frac{1}{\sqrt{2}} M_{1Z} - \frac{1}{\sqrt{3}} M_{3Z}) + \frac{1}{\sqrt{3}} \frac{V_T}{H} M_3 - \frac{2}{3} \frac{V_T^2}{\nu} M_{2ZZ} + \frac{2}{3} \frac{V_T^2}{\nu H} M_{2Z} = 0 \qquad (9)$$

$$M_{3t} - \frac{V_T}{\sqrt{3}} M_{2Z} - \frac{5}{4} \frac{V_T^2}{\nu} M_{3ZZ} + \frac{5}{4} \frac{V_T^2}{\nu H} M_{3Z} = 0$$

System (9) is the system of equation of a non-ideal liquid, the linearized system of Navier-Stokes equations with right Prandtl number for noble gases $\Pr = 2/3$. In the higher orders of the theory from the system (6) the linearized Burnett's and super-Burnett equations follow.

## 4. Construction of solutions of the fluid dynamics system by WKB method

In this section we apply the method WKB to the system(6). We shall assume, that on the bottom boundary at $z = 0$ a wave with characteristic frequency $\omega_0$ is generated. Next we choose the frequency $\omega_0$ to be large enough, to put characteristic parameter $\zeta = 3\omega_0 H / V_T \gg 1$. We shall search for the solution in the form:

$$M_N = \psi_N \exp(i\omega_0 t) + c.c. \qquad (10)$$

where, for example, $\psi_1$ corresponding to the moment $M_1$, is given by the expansion:

$$\psi_1 = \sum_{K=1}^{6} \sum_{M=1}^{\infty} \frac{1}{(i\zeta)^M} A_M^{(K)} \exp(i\zeta \varphi_K(z)), \qquad (11)$$

here $\varphi_K(z)$ - the phase functions corresponding to different roots of dispersion relation. For other moments $M_{:N}$, N=2...6 corresponding functions $\varphi_N$ are given by similar to (11) expansion. The appropriate coefficients of the series we shall designate by corresponding $B_M^{(K)}, C_M^{(K)}, D_M^{(K)}, E_M^{(K)}, F_M^{(K)}$. Substituting the series (11) in the system (6) one arrives at algebraic equations for the coefficients of (11) in each order. The condition of solutions existence results in the mentioned dispersion relation:

$$\frac{54}{125}\eta^3 + (-\frac{54}{25}iu - \frac{63}{25} + \frac{3}{5}u^2)\eta^2 + (-\frac{2}{3}iu^3 + \frac{164}{25}iu - \frac{58}{15}u^2 + \frac{18}{5})\eta - 1 - \frac{8}{3}iu + \frac{7}{3}u^2 + \frac{2}{3}iu^3 = 0 \quad (12)$$

Here for convenience the following designations are introduced:

$$(\frac{\partial \varphi_K}{\partial \bar{z}})^2 = \frac{2}{15}\eta_K, \quad u = \frac{v_0}{\omega_0}\exp(-\bar{z}), \text{ where } \bar{z} = z/H.$$

For the coefficients $A_1^{(K)}, B_1^{(K)}$ ... the algebraic relations are obtained:

$$B_1^{(K)} = -\frac{\sqrt{15}}{3}\frac{A_1^{(K)}}{\sqrt{\eta_K}}, \quad C_1^{(K)} = -\frac{A_1^{(K)}}{\sqrt{6}}\frac{(9\eta_K + 40iu - 75)}{(27\eta_K + 20iu - 30)},$$

$$C_1^{(K)} = -\frac{5}{2}\frac{\sqrt{3}}{\sqrt{\eta_K}}A_1^{(K)}\frac{(6\eta_K + 1)}{(27\eta_K + 20iu - 30)},$$

$$E_1^{(K)} = \frac{5}{\sqrt{3}}A_1^{(K)}\frac{(-30\eta_K - 10iu + 15 + 9\eta_K^2 + 10iu\eta_K)}{(27\eta_K + 20iu - 30)},$$

$$F_1^{(K)} = \frac{5}{9}A_1^{(K)}\frac{(216\eta_K + 125iu - 75 - 81\eta_K^2 - 240iu\eta_K + 50u^2 + 45iu\eta_K^2 - 50u^2\eta_K)}{(27\eta_K + 20iu - 30)\eta_K^{3/2}}$$

The dispersion relation (12) represents the cubic equation with variable coefficients, therefore the exact analytical solution by formula Cardano looks very bulky and inconvenient for analysis. We study the behavior of solutions at $v \to 0$ (free molecular regime) and $v \to \infty$ (a hydrodynamical regime).

At the limit of collisionless gas $v \to 0$ the dispersion relation becomes:

$$\frac{54}{125}\eta^3 - \frac{63}{25}\eta^2 + \frac{18}{5}\eta - 1 = 0$$

The roots are:

$\eta_1 \approx 0.37, \quad \eta_2 \approx 3.80, \quad \eta_3 \approx 1.67$

Specifying roots (12) by the theory of perturbations up to $u^3$ for the three solutions branches it is obtained:

$\eta_1 \approx 0.37 + 0.29iu + 0.14u^2 - 0.07iu^3$

$\eta_2 \approx 3.80 + 2.82iu - 1.57u^2 - 0.55iu^3$

$\eta_3 \approx 1.67 + 1.89iu + 0.04u^2 + 0.62iu^3$

Accordingly, for the $k_{I,\pm} \approx \pm\sqrt{\eta_I}$ we have:

$$k_{1,+} \approx 0.60 + 0.24iu + 0.16u^2 - 0.12iu^3$$
$$k_{2,+} \approx 1.95 + 0.72iu - 0.27u^2 - 0.04iu^3$$
$$k_{3,+} \approx 1.29 + 0.73iu + 0.22u^2 + 0.11iu^3$$

In a limit $v \to \infty$ (a hydrodynamical limit) for specifying roots(12) by the theory of perturbations up to $u^3$ for the three solutions branches it is obtained:

$$\eta_1 \approx 1.00 + 1.40iu^{-1} - 3.00u^{-2} - 6.30iu^{-3}$$
$$\eta_2 \approx 2.33 + 1.11iu - 0.20iu^{-1} + 1.16u^{-2}$$
$$\eta_3 \approx -1.39u^2 + 3.89iu + 2.50 - 1.20iu^{-1}$$

The first root relates to the acoustic branch. Accordingly, for the $k_{I,\pm} \approx \pm\sqrt{\eta_I}$ we have:

$$k_{1,+} \approx 1.00 + 0.70iu^{-1} - 1.25u^{-2} - 2.27iu^{-3}$$
$$k_{2,+} \approx \sqrt{u}(1+i)(0.75 + 0.34u^{-2}) + \sqrt{u}(1-i)(0.78u^{-1} + 0.028u^{-3})$$
$$k_{3,+} \approx 1.18iu + 1.65 + 0.09iu^{-1} - 0.64u^{-2}$$

The solution of the equation (12) at any $u$ is evaluated numerically. As an illustration let us consider a problem of generation and propagation of a gas disturbance, by a plane oscillating with a given frequency $\omega_0$. We restrict ourselves by the case of homogeneous gas, because it is the only case of existing experimental realization. We evaluate numerically the propagation velocity and attenuation factor of a linear sound.

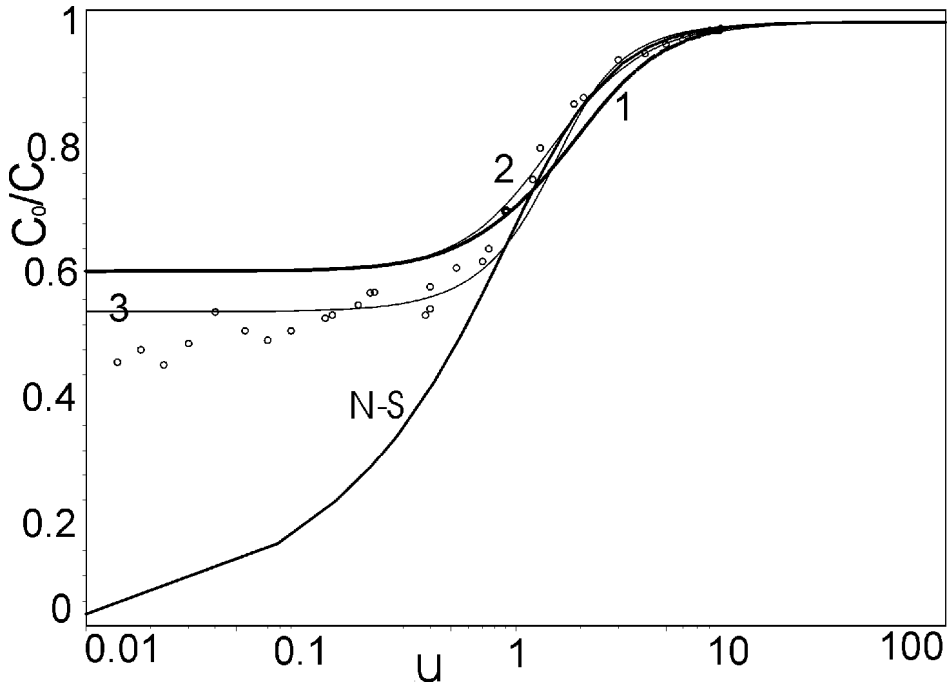

*Fig.1. The inverse non-dimensional phase velocity as a function of the inverse Knudsen number. The results of this paper for Gross-Jackson model-1 are compared to BGK model-2, Navier-Stokes, previous our work [7]-3 and the experimental data [12,13] -circle.*

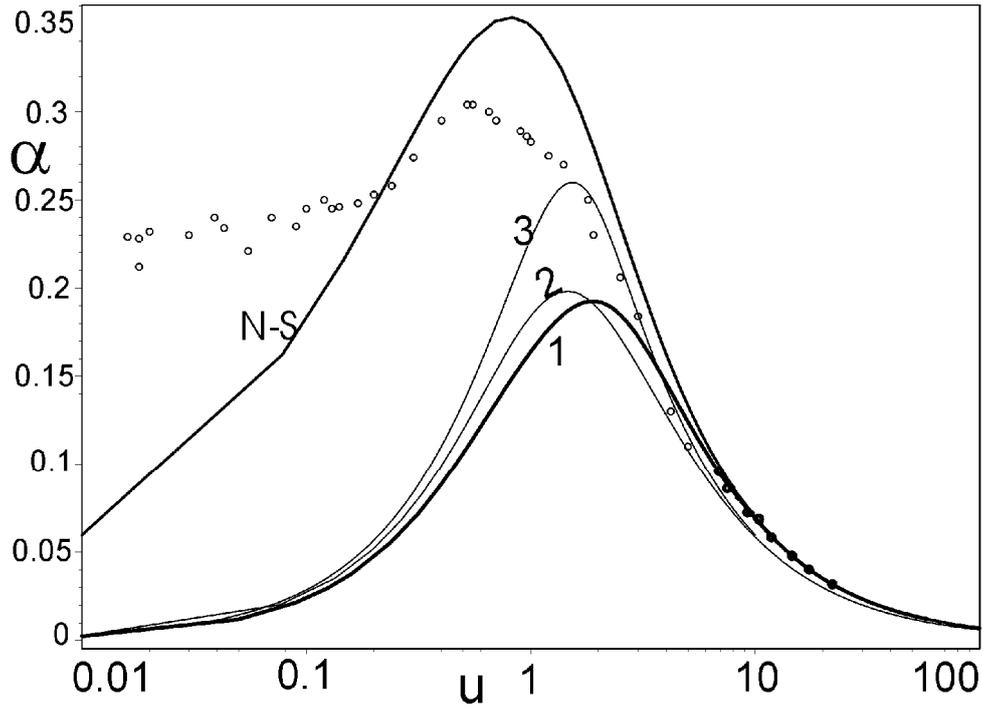

*Fig. 2. The attenuation factor of the linear disturbance as a function of the inverse Knudsen number.*

The kinetic theory[15,17] gives the good agreement with the experimental data at arbitrary Kn numbers. The Navier–Stokes prediction is qualitatively wrong at big Knudsen number. Our results for phase speed give the good consistency with the experiments at all Knudsen numbers. However, our results for the attenuation of ultrasound are good (as we can see in experiment) only for the number $r$ up to order unity. But our results look a bit better than Navier–Stokes, Alexeev [4], Chen–Rao–Spiegel[2,3] and regularization of Grad's method[18]. Unlike BGK model using the Gross-Jackson one gives right coefficients of viscosity and heat conductivity.

## 5. Conclusions

In this paper we propose a one-dimensional theory of linear disturbances in a gas, stratified in gravity field, hence propagating through regions with crucially different Kn numbers. The regime of the propagation dramatically changes from a typically hydrodynamic to the free-molecular one. We also studied three-dimensional case [10,11]. Generally the theory is based on Gross-Jackson kinetic equation, which solution is built by means of locally equilibrium distribution function with different local parameters for molecules moving "up" and "down". Equations for six moments yields in the closed fluid mechanics system. For the important generalizations of the foundation of such theory see the recent review of Alexeev [4].

## Acknowledgements

We would like to thank Vereshchagin D.A. for important discussions.

## References

1. Vereshchagin D.A., Leble S.B.. Piecewise continuous partition function and acoustics in stratified gas. Nonlinear Acoustics in Perspective, ed. R.Wei. 142-146. (1996)
2. Chen X., Rao H., and Spiegel E.~A., "Continuum description of rarefied gas dynamics: II. The propagation of ultrasound," Phys. Rev. E, **64**, 046309, (2001).
3. Spiegel E.A. and Thiffeault J.-L., Higher-order Continuum Approximation for Rarefied Gases, Physics of Fluids, **15** (11), 3558-3567. (2003)
4. Alexeev B.V., Generalized Boltzmann Physical Kinetics, (Elsevier, 2004).
5. Curtiss C.F., The classical Boltzmann equation of a gas of diatomic molecules, J. Chem. Phys. **75** 376-378. (1981)
6. Chetverushkin B.N. Kinetic schemes and quasigasdynamic system (MAKS Press, Moscow, 2004)
7. Vereshchagin D.A., Leble S.B., Solovchuk M.A.. Piecewise continuous distribution function method in the theory of wave perturbances of inhomogeneous gas. Physics Letters A, **348**, 326-334. (2006)
8. Lees L., "Kinetic theory description of rarefied gas flow" J.Soc.Industr. and Appl.Math., **13**. N 1. 278-311.( 1965)
9. Vereshchagin D.A., Leble S.B., Piecewise continuous distribution function method: Fluid equations and wave disturbances at stratified gas , physics/0503233,( 2005).
10. Solovchuk M.A., Leble S.B. The kinetic description of ultrasound propagation in a rarefied gas: from a piecewise continuous distribution to fluid equations. Proc. of Intern. Conf. "Forum Acusticum 2005" , L235-L240,( 2005).
11. Leble S.B., Solovchuk M.A. "Three-dimensional fluid equations from distribution function with discontinuity in velocity space", Mathematical modeling, **18** ,N 4, P. 118-128. (2006)
12. Meyer E., Sessler G., Z.Physik. Schallausbreirung in Gasen bei hohen Frequenzen und sehr niedriegen Drucken. **149**. P.15-39.( 1957)
13. Schotter R., "Rarefied gas acoustics in the noble gases", Phys. Fluids **8**, 1163–1168 (1974)
14. Gross E.P., Jackson E.A., "Kinetic models and linearized Boltzmann equation", Phys. Fluids **2,** N 4, 432-441 (1959).
15. Marques W. Jr., ''Dispersion and absorption of sound in monatomic gases: An extended kinetic description,'' J. Acoust. Soc. Am. **106**,3282-3288 (1999).
16. Eu B. C., and Ohr Y. G., "Generalized hydrodynamics, bulk viscosity, and sound wave absorption and dispersion in dilute rigid molecular gases." Phys. Fluids **13**, 744–753. (2001)
17. Hadjiconstantinou N. G., Garcia A. L., "Molecular simulations of sound wave propagation in simple gases." Phys.Fluids **13**, N 4, 2668-2680 (2001).
18. Struchtrup H., Torrilhon M., "Regularization of Grad's 13 moment equations: Derivation and linear analysis." Phys.Fluids **15**, N 9, 2668-2680 (2003)
19. Sharipov F., Marques W., Jr., and Kremer G. M., "Free molecular sound propagation" J. Acoust. Soc. Am. **112** (2), 395-401, (2002)